# Interface Self-Referenced Dynamic Full-Field Optical Coherence Tomography

TUAL MONFORT[1, 2], SALVATORE AZZOLLINI[1], TASNIM BEN YACOUB[1], ISABELLE AUDO[1], SACHA REICHMAN[1], KATE GRIEVE[1, 2, 4] AND OLIVIER THOUVENIN[3,4*]

[1] *Sorbonne Université, INSERM, CNRS, Institut de la Vision, 17 rue Moreau, F-75012 Paris, France*
[2] *CHNO des Quinze-Vingts, INSERM-DGOS CIC 1423, 28 rue de Charenton, F-75012 Paris, France*
[3] *Institut Langevin, ESPCI Paris, Université PSL, CNRS, 75005 Paris, France*
[4] *These authors contributed equally*
*olivier.thouvenin@espci.fr*

**Abstract:** Dynamic full-field optical coherence tomography (D-FFOCT) has recently emerged as an invaluable live label-free and non-invasive imaging modality able to image subcellular biological structures and their metabolic activity within complex 3D samples. However, D-FFOCT suffers from fringe artefacts when imaging nearby reflective surfaces and is highly sensitive to vibrations. Here, we present interface Self-Referenced (iSR) D-FFOCT, an alternative configuration to D-FFOCT that takes advantage of the presence of the sample coverslip in between the sample and the objective by using it as a defocused reference arm, thus avoiding the aforementioned artefacts. We demonstrate the ability of iSR D-FFOCT to image 2D fibroblast cell cultures, which are among the flattest mammalian cells.

## 1. Introduction

In recent years, full-field optical coherence tomography (FFOCT) has emerged as a versatile non-invasive label free optical imaging technique thanks to its high resolution, amplitude and phase contrasts [1-3], its sectioning ability, and its sensitivity and imaging speed [1-4]. Its use has been demonstrated *in vivo* on living rodent for single myelin resolution sheath disruption involved in neuropathies [5], as well as *in vivo* on human retina [6] and cornea [7] with cellular-resolution capabilities.

One promising development of FFOCT is dynamic FFOCT (D-FFOCT), in which the temporal evolution of FFOCT signal is analysed in order to quantify the nanometric active displacements of subcellular organelles [8, 9]. D-FFOCT provides a metabolic contrast [1, 9, 10] highly complementary to the structural contrast obtained from static FFOCT [11]. Static and dynamic FFOCT ((D)-FFOCT) have been combined for several *in vitro* and *ex vivo* studies, for example in retinal explants [10, 12, 13] and retinal organoids [9, 13]. Taking advantage of morphology and specific dynamic contrast, (D)-FFOCT could resolve different cell types and subcellular structures [1, 9, 10, 13-15], and monitor different cell metabolic states, such as senescence and mitosis [10, 13, 15]. As a result, (D)-FFOCT is an appealing solution to drive biology research on unaltered systems at high resolution under live imaging conditions [9, 13, 15].

Despite its success, a few aspects still impede (D)-FFOCT. Indeed, based on a Linnik interferometer configuration, (D)-FFOCT relies on two symmetric but physically separated optical arms. This aspect leads to three main drawbacks; First, (D)-FFOCT is prone to fringe artefacts when imaging close to the reflective surface of sample holders for *ex vivo* and *in vitro* studies [10, 14-17] typically preventing imaging of the first micrometers of a sample, which especially impacts the imaging of thin samples. Yet, adherent (2D) cell culture is the main *in vitro* condition used in biology [18-20]. Groux *et al.* were recently able to partially suppress these fringe artefacts by using weakly scattering porous polycarbonate membranes to move the sample away from the sample holder in 2D retina pigmented epithelial (RPE) cell cultures [15]. Nonetheless, for thinner epithelial cells (<5µm thick), such as fibroblasts, the signal from the culture membranes was partially covering the cell bodies and decreased image quality. Also, these membranes are incompatible with the highest numerical aperture (NA) objectives due to their limited working distance, and are not adaptable to all cell culture conditions. Second, D-FFOCT is sensitive to subnanometric external vibrations which, even on specifically designed vibration-free optical tables, are hard to cancel completely. This reduces the image quality of D-FFOCT and makes it difficult to properly quantify the organelle motion and the metabolic contrast [10]. Finally, (D)-FFOCT is challenging to implement using objectives with very high numerical aperture (>1), or when the interference arms contain several optical elements [13]. Indeed, any dissymmetry, including spherical aberration due to small misalignments between interfering fields results in distortions, a loss of accessible field of view, and a loss of interference contrast [13]. This typically results in non-homogeneous and distorted (D)-FFOCT signal over the field of view, which makes efficient mosaicking complicated to implement [13].

In this work we present and characterize a new interferometric configuration which overcomes these (D)-FFOCT drawbacks in order to enable imaging of cells in the vicinity of the culture surface. This *interface Self-Referenced* (iSR) (D)-FFOCT method allowed imaging of 2D flat fibroblast samples, which are of interest to biologists in disease modelling applications of mitochondrial disease [21-24]. The technique could be used more widely for rapid and robust diagnosis directly from cell phenotypes.

## 2. Methods
### 2.1 FFOCT and iSR FFOCT microscopes

FFOCT measurements were performed using the setup described in [13], and shown in a simplified form in Fig. 1. A high-power LED $S_1$ (either a M810L3 or M730L4, Thorlabs, Newport, NJ, USA, $\lambda_0$= 810 nm $\Delta\lambda$=25 nm, coherence length Lc = 8.7 µm or $\lambda_0$= 730 nm $\Delta\lambda$=40 nm, coherence length Lc = 4.4 µm), was used to illuminate a Linnik interferometer with high NA objectives (NA= 1.05, immersion medium n=1.4, silicon oil, UPLSAPO30XSIR, Olympus, Japan). The light reflected by the reference mirror, the light reflected by the glass coverslip supporting the sample, and the light backscattered by the sample are recombined by the non-polarizing beamsplitter, focused by a tube lens L3 (AC254-300-B-ML, Thorlabs, Newport, NJ, USA) to overlap, and potentially interfere, on a high full well capacity (FWC) 2D CMOS sensor (Q-2HFW, Adimec, Netherland). The lateral and axial magnifications of the system are respectively: $\gamma_t = 58, \gamma_L = 2400$.

iSR FFOCT measurements were performed on the same setup by manually blocking the reference arm, hence allowing detection of the interference between the light reflected from the glass coverslip (standard coverslips, 170 µm thick, n=1.52, P24-1.5H-N, Cellvis, Canada) and the light backscattered by the different depths of the sample. The top surface of the glass coverslip here acts as the reference mirror and is slightly defocused compared to the focal plane of the objective. The reflectivity of the glass in water is about 0.4%, according to Fresnel coefficients and neglecting angular effects.

For both configurations, the Q-2HFW camera was configured to have a FWC of 1.6 Me$^-$, with a signal to noise ratio (SNR) of 60.6 dB at saturation. Acquisitions were performed at 100 frames per second (FPS) when generating dynamic images. All samples were imaged inside a top-stage microincubator (H201-K-FRAME, H201-MW-HOLDER and OBJ-COLLAR-2532, Okolab, Italy) and were maintained at 37 C° and 5% of $CO_2$ concentration during the acquisition. Data were acquired with a custom Matlab graphical user interface enabling continuous data logging while post-processing and saving the resulting metrics in parallel threads for maximal acquisition speed [13].

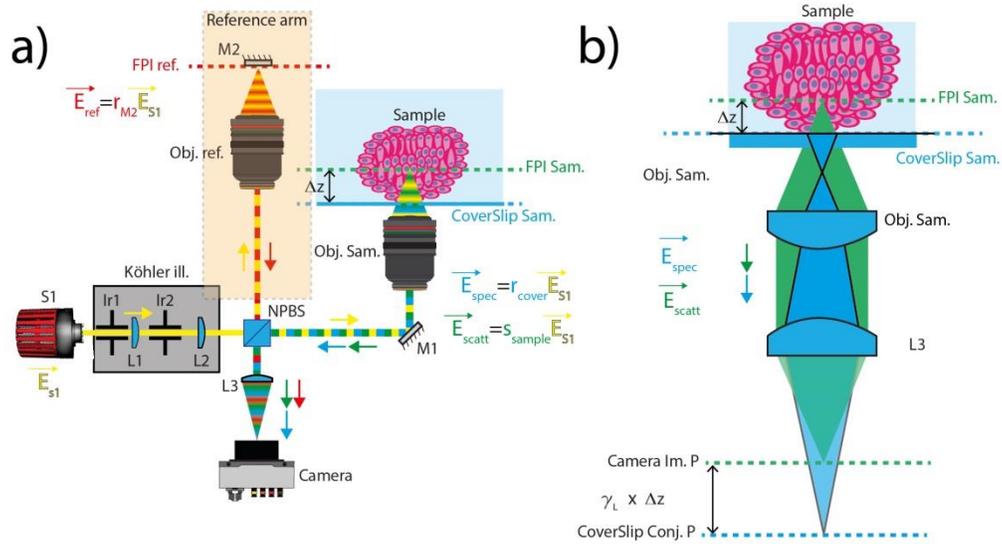

**Figure 1: Comparison between D-FFOCT setup versus iSR D-FFOCT.** Fig.1a illustrates a classical D-FFOCT set-up with a spatially incoherent source illuminating a Linnik interferometer using a Köhler illumination. The incoming field, in yellow, is split by a non-polarizer beam splitter (NPBS, BS014, Thorlabs, Newport, NJ, USA) cube into a *reference arm* and a *sample arm*. In the reference arm, an objective focuses the light on a mirror, placed at the image focal plane (FPI ref) of the reference objectif (Obj.ref). The back reflected field is sketched in red. In the sample arm, an identical microscope objective (Obj. Sam) focuses the light onto a sample, laid on a coverslip (CoverSlip Sam) for the inverted microscope, at its image focal plane (FPI Sam). The backscattered light is illustrated in green. The objective (Obj. Sam) also collects the out-of-focus light (pictured in blue) reflected by the specular top surface of the coverslip (CoverSlip Sam). Both beams are recombined by the NPBS and focussed on a camera by intermediary of a tube lens L3. In iSR FFOCT, the reference arm is blocked so that only the two beams from the sample arm reach the camera and can interfere. Fig.1b illustrates the iSR D-FFOCT configuration showing the imperfect overlap of these two beams occurring on the camera.

*2.2 Image acquisition protocol and dynamic FFOCT image generation*

The dynamic images showcased were computed according to three metrics established by Scholler *et al.* in 2020 [9] using either 512 raw FFOCT or iSR FFOCT images and computing the averaged running standard deviation with a window of 50 images, the mean frequency of the power density spectrum, and the standard deviation frequency of the power density spectrum, and displayed together in a Hue Saturation and Brightness (HSB) base, respectively.

*2.3 Fibroblast culture protocol*

Human dermal fibroblasts from a healthy subject (male, 49 y.o, Caucasian) were obtained at the Quinze-Vingts hospital after a skin biopsy. Fibroblasts were cultured in T75 flasks in DMEM glutamax medium (Thermo Fisher Scientific, 61965026) supplemented with 10% Fetal Bovine Serum (FBS), 2% sodium pyruvate, 1% penicillin-streptomycin and 1% amphotericin B, and incubated at 37°C, in a 5% CO2 incubator. When fibroblasts reached 70% confluency, cells were dissociated using Enzyme Express (1X) TrypLE™ (Thermo Fisher Scientific, 12605010). Cells were plated in an uncoated 6-well glass bottom plate (Cellvis, P06-1.5H_N) at a density of 40 000 cells/well and the medium was changed every 2 days. Five days after seeding, fibroblasts were imaged.

## 3. Results and discussion

3.1 *System characterization: Resolution, and sensitivity.*

With iSR FFOCT, we successfully obtained static images of fibroblasts by recording and subtracting two successive planes so that a phase difference of $\frac{\pi}{2}$ is obtained (not shown). We also obtained dynamic images of fibroblasts (see Fig.2-5). Interestingly, in iSR D-FFOCT fibroblasts show many small, distinguishable, highly contrasted features, such as filopodia—cylindrical structures of around 20 to 200 nm diameter [26], below the diffraction limit—that can be used to characterize the optical system. For all characterizations in subsections 3.1 and 3.2, we only use and display the Brightness channel of the D-FFOCT or iSR D-FFOCT images.

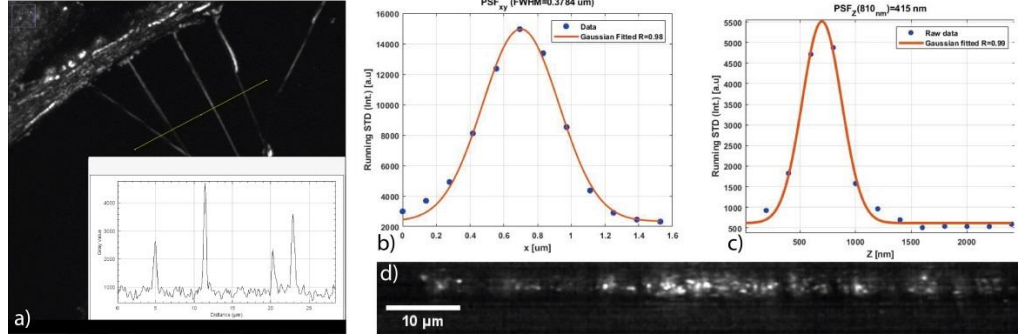

**Figure 2: Evaluation of iSR D-FFOCT spatial resolution using filipodia**, which are sub-diffraction limit sized structures. Fig.2a shows an image including filipodia from which an intensity profile is displayed and fitted to a Gaussian in Fig.2b, with a half-width at the half-maximum (HWHM) of 378.4 nm. Fig.2d. shows an axial reslicing of a z stack including filipodia from which an intensity profile is displayed and fitted to a Gaussian in Fig.2c, with a half-width at the half-maximum (HWHM) of 415 nm.

Evaluation of the system transverse and axial resolution were performed on such filipodia—see Fig.2. We obtained a very good gaussian fit (R>0.98) for both PSFs at 600 nm of depth, and measured PSFx,y ($\lambda$= 810 nm) = 378.4 nm and PSFz ($\lambda$= 810 nm) = 415 nm close to the theoretical values for the transverse resolution established for FFOCT [25], at high NA (NA>0.6), limited by diffraction [25]:

$$PSF_{theo.,x,y} = \frac{\lambda}{2NA} = 385.7 \; nm$$

$$PSF_{theo.,z}\,(NA > 0.6,) = \frac{0.44\lambda}{n(1 - cos(asin(NA/n)))} = 751.9 \; nm$$

We found that iSR D-FFOCT has a higher axial resolution than the expected theoretical resolution for FFOCT. However, we note that no theoretical model has been established for D-FFOCT in general and no PSF evaluation was previously measured in the literature, to the best of our knowledge. However, because D-FFOCT contrast relies on the non-linear post-processing of multiple images, we expect that D-FFOCT resolution may depend on the SNR of the detection and on the stochastic diffusion of the scatterer. As a result, in favorable conditions, it can exceed the resolution of FF-OCT.

iSR FFOCT can be described as a standard FFOCT system, with a defocused reference arm. Interestingly, it was recently shown that FFOCT resolution was almost insensitive to defocus [27], so that the effect of the defocus mostly results in a loss of signal.

In fibroblasts, we obtained a maximal sensitivity of 43.52 dB. Compared to standard D-FFOCT, the reference mirror in iSR D-FFOCT is made of glass of about 0.4% reflectivity, which can increase D-FFOCT sensitivity as long as incoherent reflections are lower than this value, and if the camera FWC can be saturated. Here, we used a NPBS to combine iSR D-FFOCT and a standard D-FFOCT system. However, incoherent reflections coming from the NPBS cube were very important when using a specular reflection glass/sample for creating a reference field, and still filled up the FWC of our camera: about 92% of the dynamic range of our camera is coming from incoherent reflection, of which 99% originates from the NPBS. These values were established by blocking out iteratively optics and/or sample contribution. We note further that the FFOCT configuration used in this paper picks up less incoherent reflection from the NPBS than more traditional setup [13]. Because iSR D-FFOCT is immune to mechanical vibrations, a pellicle beamsplitter could be used instead to increase the theoretical sensitivity by a factor of 12. However, more powerful light source would be required and its integration to a FF-OCT microscope would be more complex.

3.2 *System advantages compared to D-FFOCT: fringe artefacts, vibration sensitivity and mosaicking*

We first start by co-characterising iSR D-FFOCT and D-FFOCT responses in the vicinity of the coverslip, on the same areas and at the same distances from the interface coverslip/sample ($\Delta Z$). Fig.3a-b show images at the same locations at $\Delta Z=0.8$ μm, illustrating that iSR D-FFOCT contrast (Fig.3b), in the vicinity of a coverslip, is free of fringe-artefacts in comparison to D-FFOCT (Fig.3a) while displaying significant structural details, including nucleus, nucleoli, mitochondria, filipods and actin filament network (see subsection 3.3). We perform an additional Z-stacks using iSR D-FFOCT and D-FFOCT on a thicker sample in order to characterise iSR D-FFOCT and D-FFOCT axial signal response. A retinal organoid is used and iSR D-FFOCT and D-FFOCT signal is recorded on one plane every 100 nm. We found that iSR D-FFOCT (Fig.3d) maintains sufficient contrast up to $\Delta Z = 3.5$ μm while D-FFOCT only becomes fringe-free from this distance (Fig.3c).

In the case of D-FFOCT (Fig.3c), we observe a strong signal (Brightness) in the vicinity of the glass coverslip due to the glass/water reflectivity being significantly higher than light scattered by the sample, before dropping to a level where the glass coverslip and sample contribution are equivalent. These two behaviours are highlighted in Fig.3c by an orange shaded area, corresponding to the depths where D-FFOCT is not able to image samples. In the case of iSR D-FFOCT, the signal is maximal at the glass coverslip interface and drops with the depths.

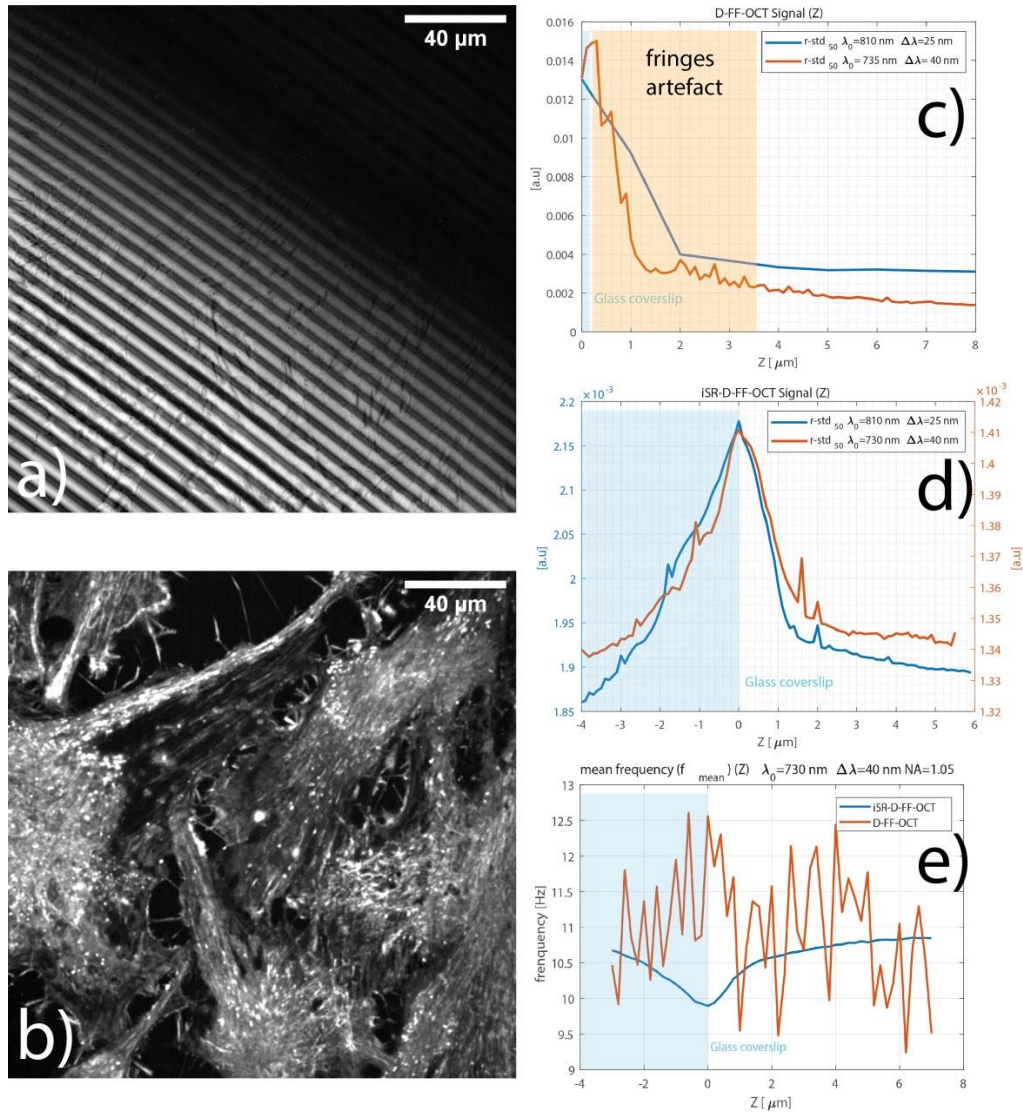

**Figure 3: Illustration of differences between D-FFOCT and iSR D-FFOCT.** Two images at the same location and depth ($\Delta Z = 0.8$ μm) are displayed using D-FFOCT (Fig.3a) and iSR D-FFOCT (Fig.3b). No fringes artefact is present when using iSR D-FFOCT (Fig.3b) unlike for D-FFOCT (Fig.3a). Signal strength (Brightness) is evaluated at different distances from the coverslip ($\Delta Z$), on a 3D sample (retinal organoid), using D-FFOCT (Fig.3c) and iSR D-FFOCT (Fig.3d). Stability of D-FFOCT and iSR D-FFOCT is assessed using the mean frequency (Hue) at different depths in a 3D sample (retinal organoid) using the LED at 730 nm.

Furthermore, we qualitatively assess the relative immunity of iSR D-FFOCT to mechanical vibration compared to D-FFOCT (Fig.3e), by measuring how the averaged mean frequency of the power spectrum density (Hue channel) evolves during a Z-stack, at the equivalent depths and location. An improvement factor of 42.08 dB is found for iSR D-FFOCT over D-FFOCT in term of mean frequency stability (Hue). Indeed, the close distance $\Delta Z$ between the scatterer and the specular reflection, used to generate a reference field, seems to correlate mechanically both. As a result, phase shifts due to vibrations are stacking up identically to the reference and scattered field. This mechanical locking result in a virtually vibration-insensitive D-FFOCT imaging modality: we did not observe artefacts while stepping around the setup nor knocking on the optical table. As a result, iSR D-FFOCT is automatically quantitative and can function without the use of a mechanically damped optical table.

Finally, iSR D-FFOCT is able to produce very homogeneous and flat interferences compared to D-FFOCT [13], thanks to its insensitivity to setup-induced aberrations, and its low penetration depth. This aspect can clearly be seen when mosaicking, as shown in Fig.4 and Fig.5a, which shows a flawless mosaic over a wide field whilst using a basic stitching method [28]. Compared to previous work using 50% overlap between D-FFOCT tiles for mosaic reconstruction at the same magnification [13], and using high end stitching method, we could use a lower overlap of 20% in a flawless manner, and we forsee lower tile overlap possible. Furthermore, because field aberrations are particularly sensitive at high NAs, and are difficult to correct, this suggests that iSR D-FFOCT could potentially be performed at higher NA than standard D-FFOCT [13], with additional elements in overlaid in the sample arm.

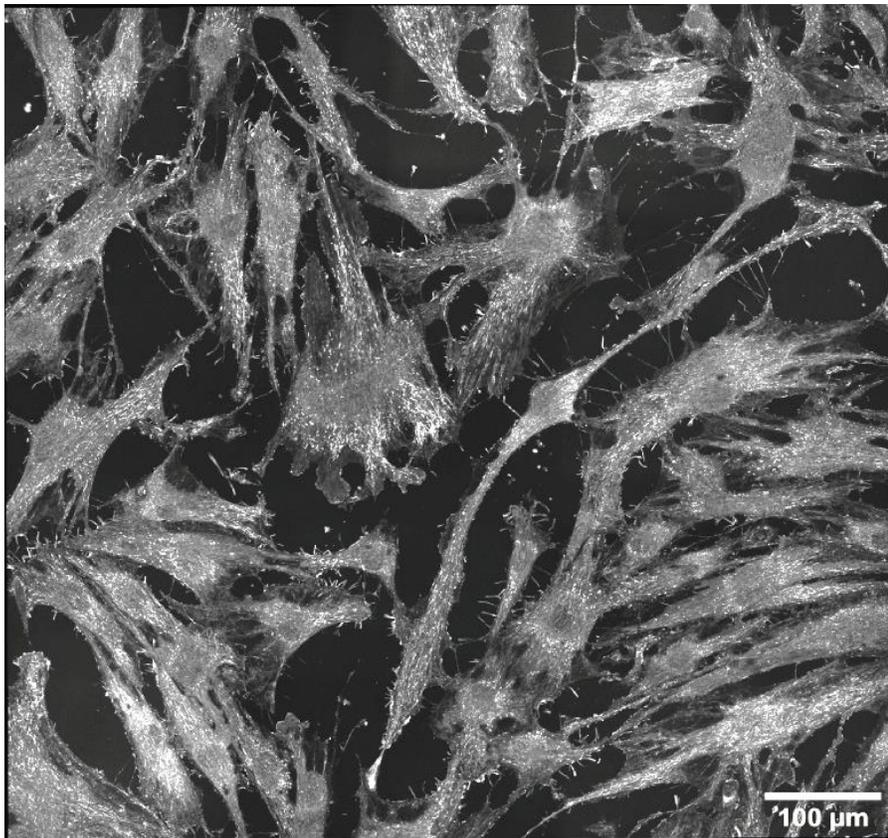

**Figure 4: iSR D-FFOCT (Brightness) mosaic of fibroblasts composed of 5x5 tiles with 20% overlap without numerical filtering.** No overlap artefact could be observed nor image field distortions confirming iSR D-FFOCT signal uniformness and flatness.

*3.3 Biological results on Fibroblasts*

We imaged fibroblasts, which are flat epithelial cells, with iSR D-FFOCT to highlight cell structures and to characterise iSR D-FFOCT imaging responses. The shape of the fibroblast can be outlined, and using volumetric acquisition, intertwined fibroblasts can be individually identified using iSR D-FFOCT's sectioning ability. Morphologically, each fibroblast displays a nucleus, pointed out by a red arrows in Fig.5a,b, d,f, with an average frequency of 5 Hz and a slight desaturation, indicating a less chaotic oscillation of the nucleus compared to its surroundings. The boundary of the nucleus could be observed at specific depth displaying a further desaturated contrast, highlighted by a dash line in Fig.5d,f. Euchromatin structures, delineated with a drop in saturation, could be observed too (Fig.5d, f). Furthermore, we observe a drop in Saturation in the cytoplasm highlighting probably vesicles, as pointed out by pink arrows in Fig.5c-d.

Bright dots, especially present in Fig.5b-c,f, are mitochondria which have been specifically identified using immunochemistry labelling and cross correlated to D-FFOCT [15]. Stress fibers linked to sites of adhesion and composed by long actin-filament bundles crossing cell bodies can also be observed as bleu filaments at 3 Hz (Fig. 5b,c, grey arrows) [29]. Moreover, filipodia which are protrusions of cytoplasm from lamellipodia, observed during cell migration, were distinguished by their high saturation and high dynamic activity at 9-13 Hz (red bundles) and by their oriented bundle of dynamic actin filaments [30]  (Fig. 5a,b,d,e, Yelow arrows). Their contrast has been specifically cross correlated with dynamic D-FFOCT contrast too [15].

Strikingly, Fig.5c-f display structures resembling to the observations made in transmission electron microscopy (TEM) by Ghilardi et al. (2021) on fibroblasts—a super resolution invasive imaging technique [21]. In particular, membrane stress in Fig.5e can be observed on the edge of the fibroblast, with a ridge-like appearance (pointed out by a white arrow). Further comparison with this work, including comparison with specific fluorescence staining images, enables identification of actin fibres (Fig.5c, purple arrow). Lamellipodia, which guide cell movement, sensing the outer environment and extracellular stimuli, are clearly visualized in our images, with pronounced broad desaturation protrusions (Fig.5b,e, blue arrows).

Overall, these results show that iSR D-FFOCT is able to detect a wide variety of cytoplasm components, cell body shape and nuclear structures in cells, showing the potential of iSR D-FFOCT for label-free and non-invasive live assessment at high resolution. Furthermore, this demonstrates that iSR D-FFOCT could be used for thin 2D cell culture imaging.

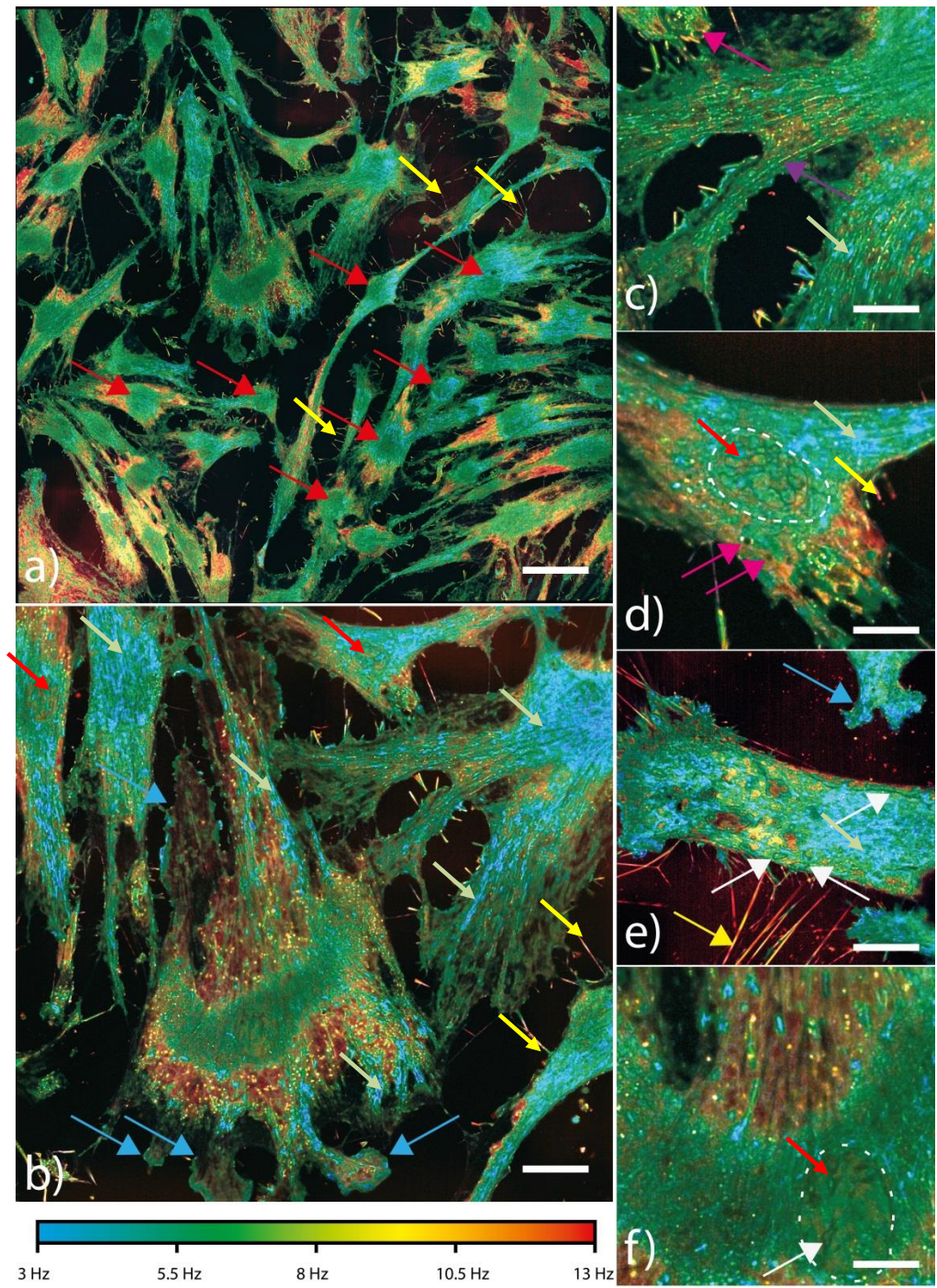

**Figure 5: Fibroblasts structures analysis using iSR D-FFOCT.** Fig.5a shows an iSR D-FFOCT HSB mosaic of fibroblasts composed of 5x5 tiles with 20% overlap. Hue scales from 3 to 13 Hz. Examples of nucleus observed are highlighted by red arrows in Fig.5a. Fig.5b highlights cases of lamellipodia structures (blue arrows). Fig.5c shows actin fibres, one example highlighted with a purple arrow. Euchromatin structures can be observed in Fig.5d within the nucleus, delimited by a white dashed line. Sharp membrane ridges

and curved membrane pockets (white arrows) are shown in Fig.5e, as well as ridges in a nucleus in Fig.5f, delimited by a white dashed line. Examples of vesicles are highlighted by pink arrows in Fig.5c-d. An Example of filipodium is highlighted in Fig.5e with a yellow arrow. Scale bar is 90 µm for Fig.5a, 40 µm for Fig.5b, 20 µm for Fig.5c, 15 µm for Fig.5d, 30 µm for Fig.5e and 15 µm for Fig.5f.

## 4. Conclusion

In this work, we have created a new label free optical microscope based on a new self-referenced configuration of *static* and *dynamic* FFOCT. This configuration is immune to fringe artefacts and mechanical vibrations. It is also less sensitive to interferometer arm misalignment and field curvature that can be critical with the use of very high NA objectives. It is simple to implement, especially on an existing D-FFOCT setup. Nonetheless, iSR D-FFOCT is restricted to an imaging depth within the coherence length of the source and is mostly interesting for thin 2D samples, as well as for recording structures at the surface of a 3D sample. These are situations where regular D-FFOCT typically struggles to obtain artefact-free images. As a result, by combining iSR D-FFOCT and D-FFOCT configurations, simply by blocking the reference arm close to the cell culture dish base, *static* and *dynamic* FFOCT images can now be obtained over a depth range from the sample surface to a few hundred microns.

Cytology plays an important role in diagnosing and managing human diseases. At a molecular level, the 'state' of a cell will depend on a large number of microscopic variables [31]. Moreover, most of the methods for measuring molecular state variables are destructive to cells, rendering it impossible to study temporal variation or correlate molecular states with downstream behaviour. In fact, organelle morphology can reflect neurological or metabolic diseases as well as cancers. For example, changes in mitochondrial morphology have been linked to several neurodegenerative disorders such as Alzheimer's [32], optic atrophy and Charcot–Marie–Tooth neuropathy [33]. Similar findings have been observed for the endoplasmic reticulum in pancreatic β-cells of type 2 diabetic patients [34]. At the nuclear level, changes in morphology have long been diagnostic of cancer with nuclear aberration correlated to the severity of prognosis [35]. A fast and easy analysis of subcellular structure with a non-invasive and live technique such as iSR D-FFOCT will contribute to enriching computational cytology and make a correct diagnosis. Moreover, this accessibility to image ultrastructure of live adherent cells will allow an efficient evaluation of new therapies targeting the pathological phenotype.

Thanks to its simplicity and its insensitivity to mechanical vibrations, iSR D-FFOCT could be deployed in much harsher environments than standard D-FFOCT, including operation rooms, classrooms, field measurements, or even within exploratory boats or zero-gravity environments, enabling the democratization of the use of D-FFOCT. In this direction, we could also imagine an adaptation of iSR D-FFOCT to cell phones, for example, based on the inexpensive extended source used for generating the flash and the camera of the mobile phone.

*Funding*

The authors thank the following sources of funding: OREO [ANR-19-CE19-0023], IHU FOReSIGHT [ANR-18-IAHU-0001], OPTORETINA (European Research Council (ERC) (#101001841)).

*Disclosures*

The authors declare that they have no known competing financial interests or personal relationships that could have appeared to influence the work reported in this paper

*Data Availability Statement*

Data and codes are available upon reasonable request from the authors.

*Acknowedgement*
We thank Jeremy Brogard and Marilou Clémençon for providing some retinal organoid.

*Author Contributions*

The overall project was imagined and conceived by TM, but supervised by OT.

The proofs of concept were made by TM.

The acquisitions were carried out by TM, with assistance of SA.

Optical design and construction were conceived and carried out by TM, supervised by OT.

Acquisition protocols were designed by TM, TBY, on samples provided by TBY, SR and IA.

Images and volumes were reconstructed by TM.

OT, TM, KG, SR and SA discussed the results and wrote the article.